\documentclass[aps,prb,reprint,twocolumn,showpacs,floatfix,superscriptaddress,nofootinbib]{revtex4}
\usepackage{graphicx}
\usepackage{amssymb}
\usepackage{amsmath}
\usepackage{lmodern}
\usepackage{color}
\usepackage{hyperref}
\usepackage{empheq}
\usepackage{epstopdf}

\begin{document}

\title{Poor man's scaling: $XYZ$ Coqblin--Schrieffer model revisited}

\author{Eugene Kogan}
\email{Eugene.Kogan@biu.ac.il}
\affiliation{Jack and Pearl Resnick Institute, Department of Physics, Bar-Ilan University, Ramat-Gan 52900, Israel}
\affiliation{Max-Planck-Institut f\"{u}r Physik komplexer Systeme,  Dresden 01187, Germany}
\author{Zheng Shi}
\email{zheng.shi@fu-berlin.de}
\affiliation{Dahlem Center for Complex Quantum Systems and Physics Department,
Freie Universitat Berlin, Arnimallee 14, 14195 Berlin, Germany}

\begin{abstract}
We derive the third-order poor man's scaling equation for a generic Hamiltonian describing a quantum impurity embedded into an itinerant electron gas. We show that the $XYZ$ Coqblin--Schrieffer model introduced by one of us earlier is algebraically renormalizable in the sense that the form of the Hamiltonian is preserved along the scaling trajectory, write down the scaling equations for the model, and analyze the renormalization group flows in the cases of both constant and pseudogap densities of states.
\end{abstract}

\maketitle

\section{Introduction}

In the celebrated Kondo problem, a seemingly innocuous magnetic impurity coupled to a band of itinerant electrons gives rise to an infrared logarithmic divergence in not only resistivity but also almost all thermodynamic and kinetic properties \cite{kondo,hewson}. To interpret this logarithmic divergence, Anderson\cite{anderson} proposed the idea of the so-called poor man's scaling: the effects of high-energy excitations can be absorbed into renormalized coupling constants at low energies. It was later realized that similar physics is found in many more complicated impurity models with internal degrees of freedom. Among these is the Coqblin--Schrieffer (CS) model motivated by the orbital degeneracy of transition metal ions with unfilled d or f shells \cite{coqblin,cox,hewson}. The CS model has recently attracted renewed interest in various contexts\cite{kikoin,desgranges,figueira,avishai}.

The study of spin anisotropy\cite{cox,costi,irkhin2,thomas} has yielded rich results as an offshoot of the original isotropic Kondo problem. Following works on the spin-anisotropic Kondo model\cite{anderson,shiba,yosida,kogan}, one of the authors introduced anisotropic CS models and derived the poor man's scaling equations for these models\cite{kogan2,kogan3}. In this work we return to the consideration of what we previously called the $XYZ$ CS model; the scaling equations are now evaluated to the third order, an error in the previously obtained second-order scaling equations\cite{kogan2} is corrected, and we explore the scaling flow diagrams in detail. We also take into account a possible power-law energy dependence of the density of states of itinerant electrons at the Fermi energy (i.e. a pseudogap density of states)\cite{fradkin,cassanello,gbi,vojtabulla,fritzvojta,mitchellfritz,kogan,shinicaaffleck}, which can arise in semimetals, nodal superconductors as well as one-dimensional interacting systems.

The rest of the paper is constructed as follows. In Section~\ref{sec:scal} we present the third-order scaling equation for the coupling constants and review the notion of algebraic renormalizability for a rather generic quantum impurity model embedded into an itinerant electron gas. In Section~\ref{sec:kondo} we consider the solutions to the poor man's scaling equations for the $XYZ$ Kondo model, which is the $N=2$ special case of the $XYZ$ CS model, and plot the corresponding three-dimensional weak-coupling flow diagrams. We first review the case of a constant density of states for the itinerant electrons, expanding on the well-known limit of the $XXZ$ Kondo model. We then turn to a pseudogap density of states, and generalize the analysis of the $XXZ$ Kondo model in Refs.~\onlinecite{kogan,kogan2} to the fully anisotropic $XYZ$ case. In Section~\ref{sec:ani} we present the poor man's scaling equations for the $XYZ$ CS model in the more general case $N>2$, again for both constant and pseudogap densities of states, and analyze the three-dimensional flow diagrams. Section~\ref{sec:conclusion} concludes the paper. Appendix~\ref{sec:appv3} describes the derivation of the third-order scaling equation for the generic quantum impurity model. In Appendix~\ref{sec:renorm} we show explicitly the algebraic renormalizability of the $XYZ$ CS model at the second order. Some additional mathematical details are relegated to Appendix~\ref{sec:pfaff}.

\section{Scaling equation and algebraic renormalizability\label{sec:scal}}

The quantum impurity that we consider is coupled to conduction electrons and described by the Hamiltonian\cite{cox,kogan2,kogan3}
\begin{eqnarray}
\label{hamilto}
H=\sum_{{\bf k}\alpha}\epsilon_{\bf k}c_{{\bf k}\alpha}^{\dagger}c_{{\bf k}\alpha}
+\sum_{\substack{{\bf k},{\bf k}'\\\alpha\beta,ab}}V_{\beta\alpha,ba}X_{ba}c_{{\bf k}'\beta}^{\dagger}c_{{\bf k}\alpha},
\end{eqnarray}
where $c^{\dagger}_{{\bf k}\alpha}$ creates a conduction electron with wave vector ${\bf k}$, channel $\alpha$, and energy $\epsilon_{\bf k}$. The Hubbard $X$-operator is defined as $X_{ba}=|b\rangle \langle a|$, where $|a\rangle,|b\rangle$ are the impurity states.

\begin{figure}
\begin{center}
\includegraphics[width=0.9\columnwidth]{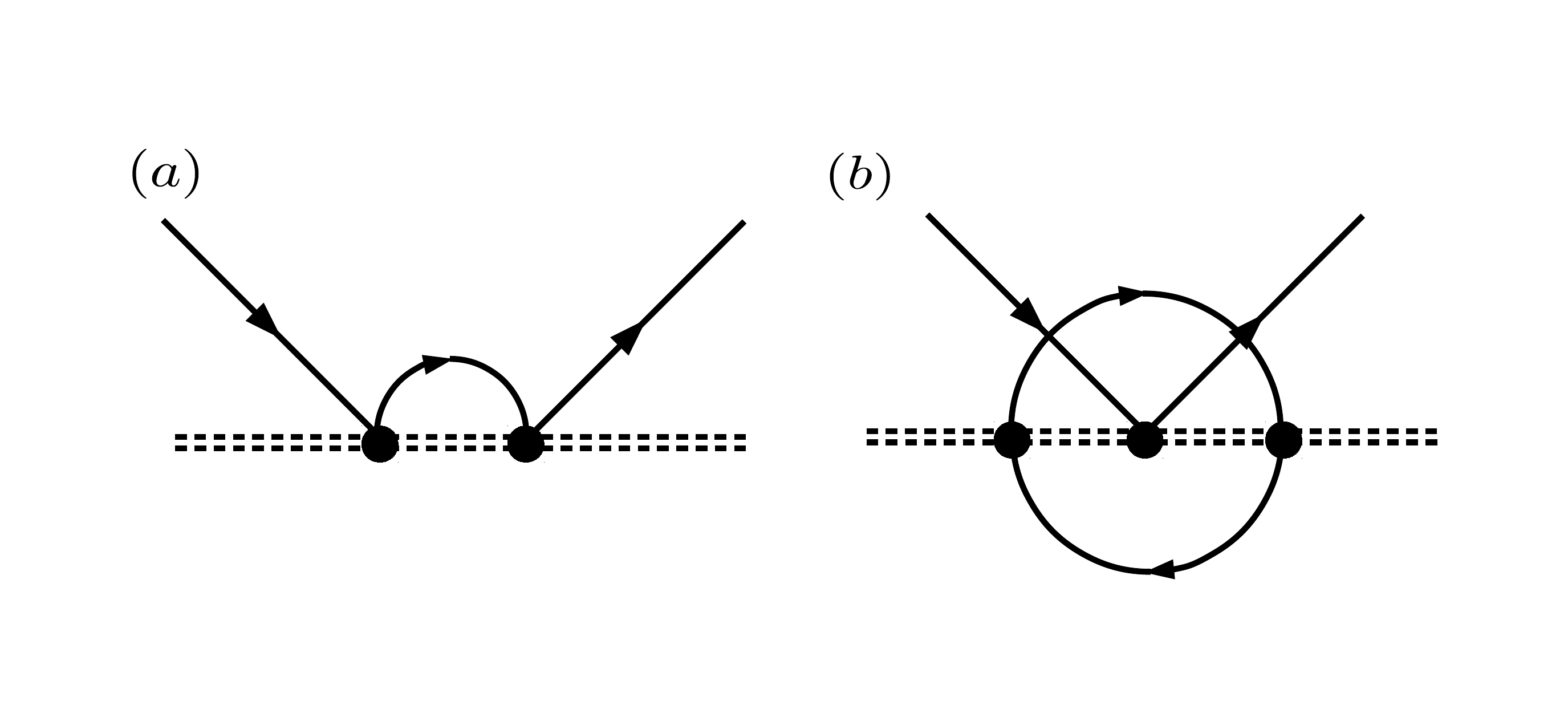}
\end{center}
\caption{Diagrams contributing to the scaling equation Eq.~(\ref{v3sceq}) at (a) the second order and (b) the third order. Solid lines and double dashed lines represent electrons and impurity propagators respectively.}\label{fig:feynman}
\end{figure}

While studying the physics in the vicinity of the Fermi energy, we must account for the virtual transitions from and to electron states at higher energies. In the poor man's scaling formalism \cite{anderson}, one reduces the semi-bandwidth of the conduction electrons from $D$ to $D-|dD|$ ($dD<0$ is infinitesimal), discarding the electronic states in the energy intervals $(D-|dD|,D)$ and $(-D,-D+|dD|)$; however, virtual transitions through these states are retained in the form of a modified coupling constant $V$, such that the impurity scattering matrix elements are the same at low energies. The coupling $V$ is therefore renormalized as the energy scale $D$ is reduced. To the order $O(V^3)$, the diagrams in Fig.~\ref{fig:feynman} produce the following scaling equation for the generic Hamiltonian Eq.~(\ref{hamilto}):
\begin{eqnarray}
\label{v3sceq}
&&\frac{dV_{\beta \alpha ,ba}}{d\ln \Lambda } =\rho \sum_{\gamma c}\left(
V_{\beta \gamma ,bc}V_{\gamma \alpha ,ca}-V_{\gamma \alpha ,bc}V_{\beta
\gamma ,ca}\right)  \notag \\
&& -\rho ^{2}\sum_{\delta \gamma }\sum_{cd}V_{\delta \gamma
,bc}V_{\beta \alpha ,cd}V_{\gamma \delta ,da}  \notag \\
&&+\frac{1}{2}\rho ^{2}\sum_{\gamma \delta cd}\left( V_{\delta \gamma
,bc}V_{\gamma \delta ,cd}V_{\beta \alpha ,da}+V_{\beta \alpha ,bc}V_{\delta
\gamma ,cd}V_{\gamma \delta ,da}\right) \text{.}  \label{3sc}
\end{eqnarray}
where $\Lambda=D/D_0$ and $D_0$ is the initial semi-bandwidth. The second-order terms in Eq.~(\ref{v3sceq}) are already given in Refs.~\onlinecite{kogan2,kogan3}; Appendix~\ref{sec:appv3} explains in detail how the third-order terms are obtained.

A particular interesting case is when $V_{\beta\alpha,ba}$ can be written as a sum of direct products of Hermitian matrices $\{G^p\}$ and $\{\Gamma^p\}$, which act respectively in impurity and channel Hilbert spaces:
\begin{eqnarray}
\label{producti}
V=2\sum_iJ_i\sum_{p\in \{P_i\}}G^p\otimes \Gamma^{\tilde{p}}.
\end{eqnarray}
Here $\{G_p\}$ ($p\in \{P_i\}$) is a given set of generators corresponding to the coupling constants $J_i$; $\Gamma^{\tilde{p}}$ in most cases considered in Ref.~\onlinecite{kogan3} is just the generator isomorphic to the $G^p$, but in general, it can be another $i$-specific generator $\Gamma$ corresponding to $G^p$. Under the assumption of Eq.~(\ref{producti}), there will be fewer coupling constants than the maximum number of entries in the matrix $V_{\beta\alpha,ba}$, and it is clear that not all interactions in the form of Eq.~(\ref{producti}) have this form preserved by scaling (or ``algebraically renormalizable'') even at the second order. The problem of finding algebraically renormalizable interactions usually requires symmetry considerations of the group corresponding to $\{G^p\}$ and $\{\Gamma^p\}$, and has been discussed in Refs.~\onlinecite{kogan2,kogan3} at the second order. In particular, the search for algebraically renormalizable models has led to the proposal of the $XYZ$ CS model, which we will focus on in the remainder of this paper:

\begin{eqnarray}
\label{cs01b}
V&=&J_S\sum_{m\neq m'}X_{mm'}c_{m'}^{\dagger}c_{m}+J_A\sum_{m\neq m'}X_{mm'}c_{m}^{\dagger}c_{m'}\nonumber\\
&+&J_z\sum_m X_{mm}c_{m}^{\dagger}c_m-\frac{J_z}{N}\sum_{mm'}X_{mm}c_{m'}^{\dagger}c_{m'}.
\end{eqnarray}
Here we have suppressed all momentum labels for clarity. Below we discuss the scaling flows of Eq.~(\ref{cs01b}) at the third order as predicted by Eq.~(\ref{3sc}). Appendix~\ref{sec:renorm} shows the detailed calculation of the second-order terms, thus demonstrating the algebraic renormalizability of Eq.~(\ref{cs01b}) explicitly at the second order; at the third order we only present the final results.

\section{$XYZ$ Kondo model\label{sec:kondo}}

\subsection{From spins to Hubbard operators}

To motivate and explain our treatment of the $XYZ$ CS model in Sec.~\ref{sec:ani}, let us start from the analysis of the spin-anisotropic Kondo model
\begin{eqnarray}
\label{hamiltonian}
H=\sum_{{\bf k}\alpha}\epsilon_{\bf k}c_{{\bf k}\alpha}^{\dagger}c_{{\bf k}\alpha}
+\sum_{{\bf k}{\bf k}'\alpha\beta} J_{ij}S^i\sigma^j_{\alpha\beta}c_{{\bf k}'\alpha}^{\dagger}c_{{\bf k}\beta},
\end{eqnarray}
where $S^x,S^y,S^z$ are the impurity spin operators, $\sigma^x,\sigma^y,\sigma^z$ are the Pauli matrices, $J_{ij}$ is the anisotropic exchange coupling matrix, and summation with respect to any repeated Cartesian index is implied. After the Hamiltonian Eq.~(\ref{hamiltonian}) is reduced to principal axes $J_{ij}=J_{i}\delta_{ij}$, the corresponding scaling equations are
\begin{eqnarray}
\label{scalingsc02}
\frac{dJ_x}{d\ln\Lambda} &=& -2J_yJ_z+J_x (J_y^2+J_z^2),\nonumber\\
%\label{scalingsc03}
\frac{dJ_y}{d\ln\Lambda} &=& -2J_xJ_z+J_y (J_x^2+J_z^2),\\
%\label{scalingsc03}
\frac{dJ_z}{d\ln\Lambda} &=& -2J_xJ_y+J_z (J_x^2+J_y^2).\nonumber
\end{eqnarray}
(Here and further on we take the constant density of states of the itinerant electrons to be equal to 1).

When we neglect the third-order terms (which is justified by the assumption of weak coupling) \cite{shiba,kogan}, the general solution of Eq.~(\ref{scalingsc02}) can be  written in terms of elliptic functions
\begin{eqnarray}
\label{amm9}
J_{\alpha} &=&A\;\mathrm{ns}(At+\psi,k)\nonumber\\
J_{\beta}&=&A\;\mathrm{cs}(At+\psi,k)\\
J_{\gamma} &=&A\;\mathrm{ds}(At+\psi,k),\nonumber
\end{eqnarray}
where $\{\alpha,\beta,\gamma\}$ is an arbitrary permutation of  $\{x,y,z\}$, and $t=2\ln\Lambda$.
In the general case ($k\neq 0,1$) the flow lines  go to infinity both with decreasing and with increasing $t$. (A flow line starting and ending at finite energies corresponds to a finite interval of $At$.)
A flow line is attracted to the asymptotic ray $J_{\alpha}=J_{\beta}=J_{\gamma}>0$ with decreasing $t$ and $J_{\alpha}=J_{\beta}=-J_{\gamma}>0$ with increasing $t$,
or to the asymptotic ray $J_{\alpha}=J_{\beta}=-J_{\gamma}>0$ with decreasing $t$
and to $J_{\alpha}=J_{\beta}=J_{\gamma}<0$ with increasing $t$.
We see that there are four strong-coupling phases in total, each one corresponding to the three-dimensional attraction region of the appropriate ray. The third-order terms do not change these conclusions, because they would only become important at strong coupling where we expect the perturbation theory to fail.

From the scaling equation Eq.~(\ref{scalingsc02}) itself follows the existence of 6 planes,
\begin{eqnarray}
\label{sp}
J_x&=&J_y,\;\;J_y=J_z,\;\;J_z=J_x\nonumber\\
J_x&=&-J_y,\;\;J_y=-J_z,\;\;J_z=-J_x,
\end{eqnarray}
each one being invariant under the scaling flow.
These planes form in some sense the skeleton of flow diagram.  From
Eq.~(\ref{amm9}) it additionally follows that  parts of these planes play the role of separatrices. Thus the phase characterized by the attractor $J_{\alpha}=J_{\beta}=J_{\gamma}>0$ is the space angle with 3 faces defined by the inequalities
$J_x+J_y>0$, $J_x+J_z>0$, $J_y+J_z>0$.
Other three phases can be obtained from that one by making rotations from the group of tetrahedron.

The flow lines on the invariant planes should be considered separately. Taking, for example, the plane $J_x=J_y$, we return to the previously well-studied $XXZ$ Kondo model,
\begin{eqnarray}
\label{lingsc01d}
\frac{dJ_x}{d\ln\Lambda}&=&-2J_xJ_z+J_x (J_x^2+J_z^2),\nonumber\\
\frac{dJ_z}{d\ln\Lambda}&=&-2 J_x^2+2J_z J_x^2.
\end{eqnarray}
These scaling equations exhibit Kosterlitz-Thouless (KT) physics in their range of validity: initial parameters satisfying $0<|J_x|<-J_z$ lead to a flow towards the fixed line $J_x=0$, $J_z<0$; otherwise, either $J_z>0$ or $0<-J_z<|J_x|$ results in a flow to strong coupling $|J_x|\to\infty$ and $J_z\to\infty$. The separatrix line between the two regimes is $|J_x|=-J_z$. Thus, considering, for example, the line $J_x=J_y=J_z$, we understand that the ray from the origin to $+\infty$
serves as an attractor, and the ray from the origin to $-\infty$  serves as the separatrix line on the invariant plane.
The invariant planes also contain the lines of fixed points, each one corresponding to two of the three $J_i$ being equal to zero. Each fixed point has a one-dimensional attraction region and hence is a critical point.

Notice that the flow diagram has a simple geometric meaning at the second order: each flow line can be considered as the intersection of two parabolic cylinders, one belonging to the family $J_x^2-J_y^2=C_1$, and the other belonging to the family $J_x^2-J_z^2=C_2$.

To motivate the consideration of the $XYZ$ Coqblin-Schrieffer model in Section~\ref{sec:ani}, we now write down the interaction in Eq.~(\ref{hamiltonian}) using Hubbard $X$-operators
\begin{eqnarray}
\label{huhu}
V&=&J_S\left(X_{+-}c_{-}^{\dagger}c_{+}+X_{-+}c_{+}^{\dagger}c_{-}\right)\nonumber\\
&+&J_A\left(X_{+-}c_{+}^{\dagger}c_{-}+X_{-+}c_{-}^{\dagger}c_{+}\right)\nonumber\\
&+&J_z\left(X_{++}c_{+}^{\dagger}c_{+}+X_{--}c_{-}^{\dagger}c_{-}\right)\nonumber\\
&-&\frac{1}{2}J_z\left(X_{++}+X_{--}\right)\left(c_{+}^{\dagger}c_{+}+c_{-}^{\dagger}c_{-}\right),
\end{eqnarray}
where $J_S=(J_x+J_y)/2$ and $J_A=(J_x-J_y)/2$. (We again omit the wave vector indices.)
It turns out that such an interaction is also algebraically renormalizable, and the scaling equation can be written down as
\begin{eqnarray}
\label{lingsc01c}
\frac{dJ_S}{d\ln\Lambda}&=&-2J_SJ_z+J_S (J_S^2-J_A^2+J_z^2),\nonumber\\
%\label{scalingsc02b}
\frac{dJ_A}{d\ln\Lambda}&=&2 J_AJ_z+J_A (-J_S^2+J_A^2+J_z^2),\\
%\label{scalingsc03b}
\frac{dJ_z}{d\ln\Lambda}&=&-2 (J_S^2-J_A^2)+2J_z (J_S^2+J_A^2).\nonumber
\end{eqnarray}
Of course, Eq.~(\ref{lingsc01c}) can also be obtained from Eq.~(\ref{scalingsc02}).

\begin{figure}
\begin{center}
\includegraphics[width=0.9\columnwidth]{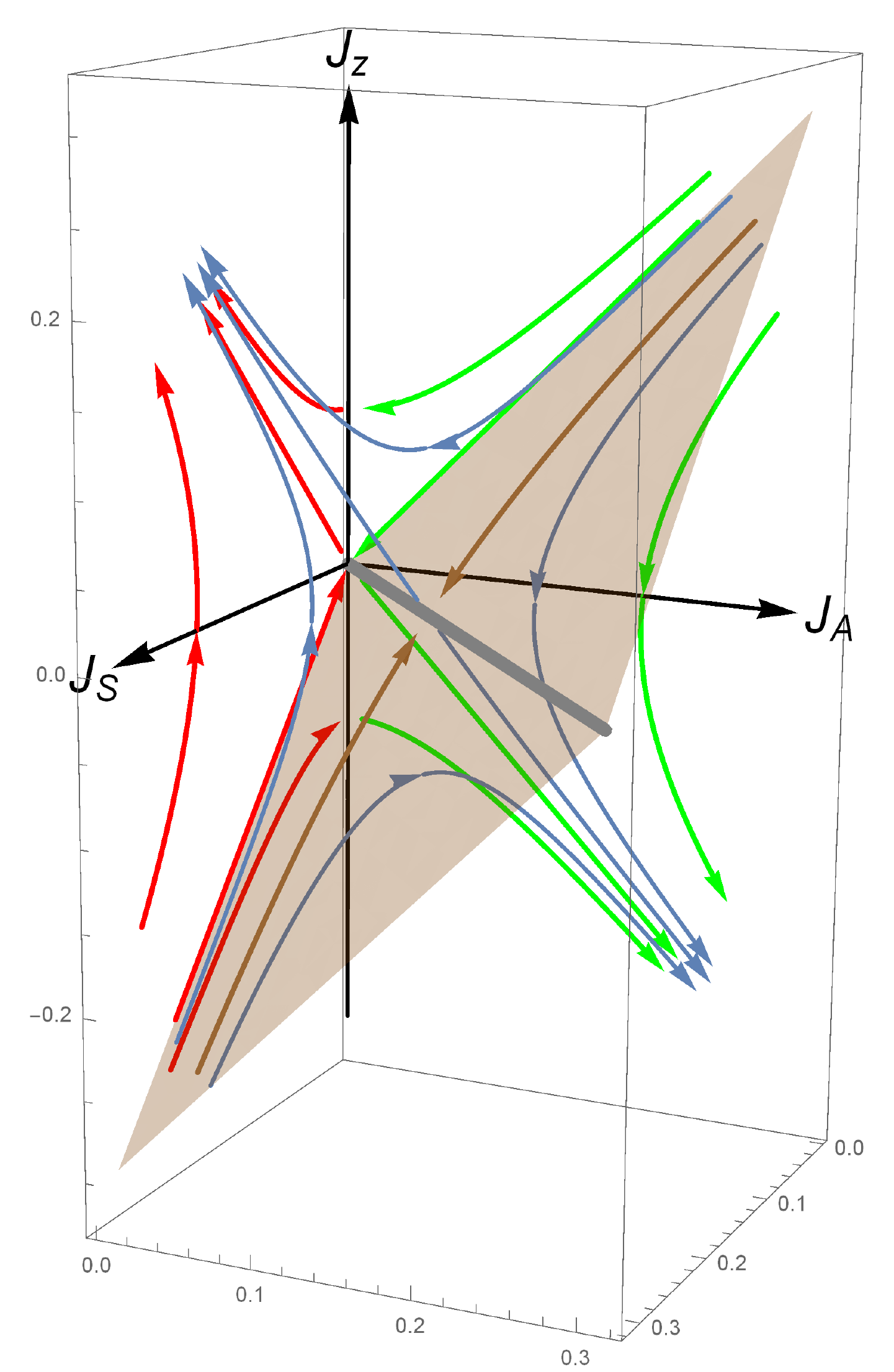}
\end{center}
\caption{Numerical three-dimensional flow diagram of the $XYZ$ Kondo model, obtained from the third-order weak-coupling scaling equation Eq.~(\ref{lingsc01c}). Only the $J_S\geq0$ and $J_A\geq 0$ part is shown; the rest of the diagram follows from the symmetries $J_S\to -J_S$ and $J_A\to-J_A$. The red trajectories lie on the $J_A=0$ plane and the green ones lie on the $J_S=0$ plane; both exhibit Kosterlitz-Thouless physics. The brown trajectories are on the light brown phase boundary $J_z+J_S-J_A=0$, and the gray line is the fixed line $J_S=J_A$ and $J_z=0$. The remaining blue trajectories are between phase boundaries. Two strong-coupling phases are shown: $J_S\to\infty$ and $J_z\to\infty$; $J_A\to\infty$ and $J_z\to-\infty$.}\label{fig:flowKondo}
\end{figure}

The statements of the present Section are illustrated by the numerical flow diagram Fig.~\ref{fig:flowKondo}, which we plot by numerically integrating the weak-coupling scaling equation.  Because of the symmetries $J_S\to -J_S$ and $J_A\to-J_A$, it suffices to focus on the case $J_S\geq0$ and $J_A\geq 0$. Scaling flows in the various limiting cases we have considered are highlighted.

\subsection{Pseudogap density of states}

The flow diagram of the $XYZ$ Kondo model is much more interesting when the itinerant electrons have a (local) density of states with a power-law dependence upon the energy because of, for instance, the electron dispersion\cite{fradkin}:
\begin{eqnarray}
\label{e}
\rho(\epsilon)=C|\epsilon|^r,\;\;\;\text{if}\;\;|\epsilon|<D.
\end{eqnarray}
This model was considered by us previously \cite{kogan,kogan2} but only the flow diagram of the $XXZ$ case ($J_x=J_y$) was discussed. In this section, we clarify the full three-dimensional flow diagram as a special case of the $XYZ$ CS model. We limit ourselves to the particle-hole symmetric case, because particle-hole symmetry breaking perturbations are known to change the phase diagram drastically in a way that is difficult to analyze using our weak-coupling method\cite{gbi}.

The scaling equation in the appropriate units (for details see Refs.~\onlinecite{kogan,kogan2}) is
\begin{eqnarray}
\label{scalinga00}
\frac{d J_x}{d\ln\Lambda}&=&rJ_x-2J_yJ_z+J_x (J_y^2+J_z^2) \nonumber\\
\frac{d J_y}{d\ln\Lambda}&=&rJ_y-2J_xJ_z+J_y (J_x^2+J_z^2) \\
\frac{d J_z}{d\ln\Lambda}&=&rJ_z-2J_xJ_y+J_z (J_x^2+J_y^2).\nonumber
\end{eqnarray}
In the weak-coupling regime Eq.~(\ref{scalinga00}) has a trivial fixed point $J_x=J_y=J_z=0$ corresponding to a decoupled impurity spin, and four nontrivial fixed points
\begin{eqnarray}
\label{odd}
\left|J_x\right|=\left|J_y\right|=\left|J_z\right|=\frac{r}{2}+O(r^2);\;\;\; \;\;\;J_xJ_yJ_z>0,
\end{eqnarray}
describing a finite isotropic Heisenberg exchange. Linear analysis in the vicinity of the fixed points shows that the trivial fixed point is stable, and hence describes the decoupled phase. Nontrivial fixed points are semistable, and hence are critical points of the model.

When we ignore the third-order terms, the general solution of Eq.~(\ref{scalinga00}) can be written in terms of elliptic functions and contains three parameters $A,\psi,k$,  \cite{kogan,kogan2}
\begin{eqnarray}
\label{amm0}
J_{\alpha} &=&A\lambda\cdot\mathrm{ns}(A\lambda+\psi,k)\nonumber\\
J_{\beta}&=&A\lambda\cdot\mathrm{cs}(A\lambda+\psi,k)\\
J_{\gamma} &=&A\lambda\cdot\mathrm{ds}(A\lambda+\psi,k),\nonumber
\end{eqnarray}
where  $\lambda=2\Lambda^r$, and $\{\alpha,\beta,\gamma\}$ is an arbitrary permutation of  $\{x,y,z\}$.

Notice that Eq.~(\ref{amm0}) describes a two-parameter family of the flow lines (the parameters being $\psi$ and $k$). Eq.~(\ref{amm9}) also describes a two-parameter family of the flow lines, but the parameters are $A$ and $k$, the former being just a trivial scale parameter.

Eq.~(\ref{amm0}) clearly shows the existence of the decoupled phase, and the strong-coupling phases, identical to those obtained in the previous Section. Eq.~(\ref{amm0}) also shows
that each nontrivial fixed point belongs to a critical surface $\psi=0$, which separates one of the strong-coupling phases from the decoupled phase.

\begin{figure}
\begin{center}
\includegraphics[width=0.9\columnwidth]{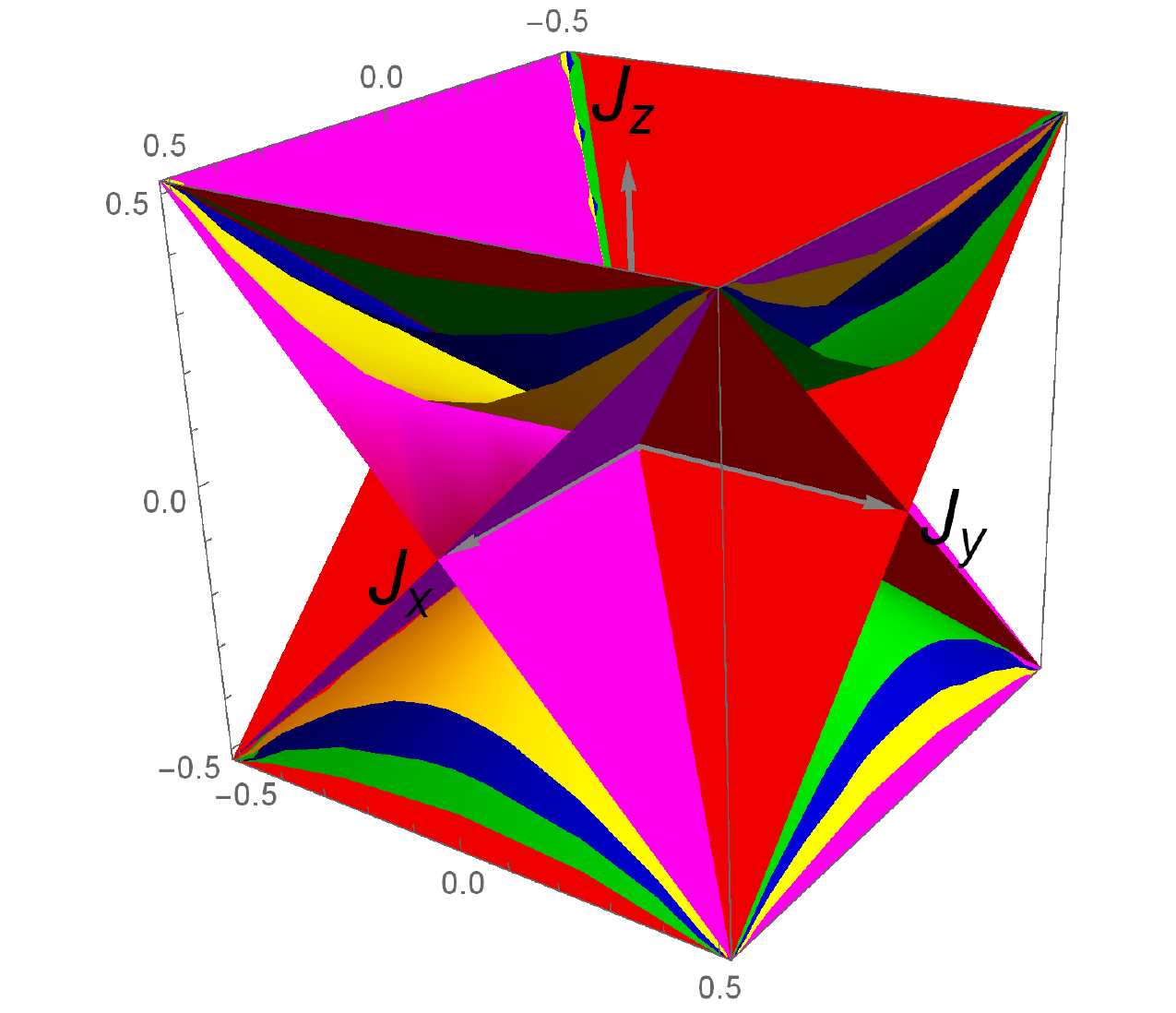}
\end{center}
\caption{Elliptic cones Eq.~(\ref{cone}) containing second-order flow trajectories of the pseudogap Kondo model. Red, green, blue, yellow and magenta surfaces have $k=0$, $k=1/2$, $k=1/\sqrt{2}$, $k=\sqrt{3}/2$ and $k=1$, respectively.\label{fig:cones}}
\end{figure}

From Eq.~(\ref{amm0}) we see that each flow line lies completely on the elliptic cone\cite{kogan,kogan2}
\begin{equation}
\label{cone}
(1-k^2) J_{\alpha}^2+k^2 J_{\beta}^2-J_{\gamma}^2=0.
\end{equation}
It should be noted that this property holds only when the third-order terms in the scaling equation are negligible. This family of elliptic cones foliates the phase space, and all touch each other along the isotropic lines $\left|J_x\right|=\left|J_y\right|=\left|J_z\right|$, as shown in Fig.~\ref{fig:cones}.

\begin{figure}
\begin{center}
\includegraphics[width=0.9\columnwidth]{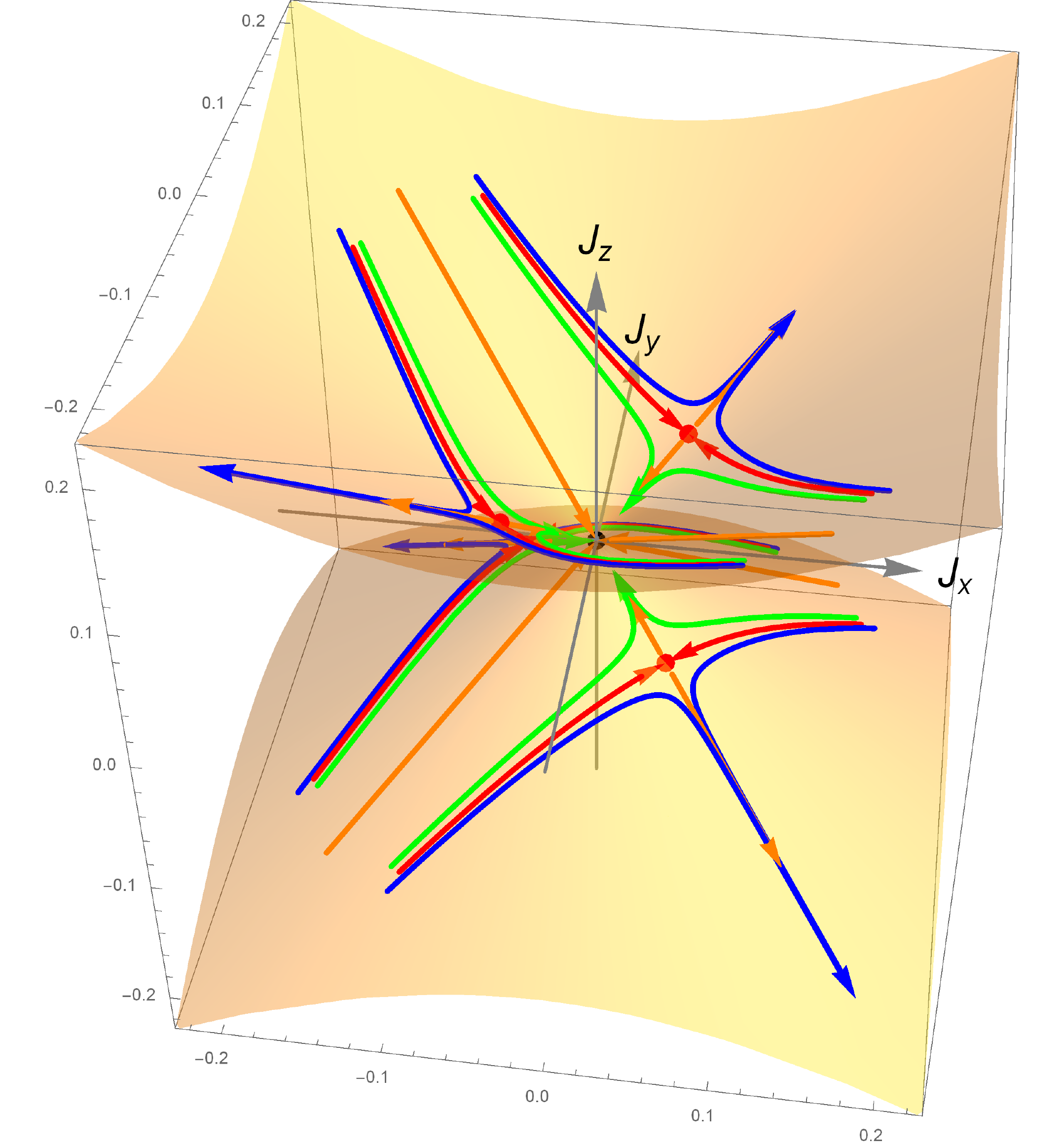}
\end{center}
\caption{Typical flow trajectories of the pseudogap $XYZ$ Kondo model with $r=0.1$ on the surface of a cone Eq.~(\ref{cone}); we have chosen $k=1/\sqrt{2}$ and ignored the third-order terms in Eq.~(\ref{scalinga00}). The trivial fixed point at the origin is painted in black, and the nontrivial fixed points in red. The isotropic orange flow trajectories are shared by all elliptic cones; the green trajectories flow towards weak coupling, the blue ones towards strong coupling, and the red trajectories lie on phase boundaries.\label{fig:onecone}}
\end{figure}

For the degenerate cases $k=0$ or $k=1$ the elliptic cone becomes a pair of planes. The flow trajectories on such planes were presented in our previous publication\cite{kogan}. Here, in Fig.~\ref{fig:onecone}, we show the trajectories on one of the nondegenerate cones, which nevertheless constitute a general representation of the flow diagram because the behaviors of different cones are rather similar. When the third-order terms are included, the flow diagram remains qualitatively the same, although we can no longer classify the trajectories by elliptic cones.

\section{$XYZ$ Coqblin-Schrieffer model\label{sec:ani}}

In this section, we turn our attention to the $XYZ$ CS model with an arbitrary number of channels $N$, thereby generalizing our $N=2$ results in Section~\ref{sec:kondo}.

\subsection{Constant density of states}

The CS model with full $SU(N)$ symmetry\cite{coqblin} is represented by the interaction
\begin{equation}
\label{cs0}
V=J\sum_{mm'} X_{mm'}c_{m'}^{\dagger}c_m  -(J/N)\sum_{mm'}X_{mm}c_{m'}^{\dagger}c_{m'},
\end{equation}
where the quantum number $m,m'=1,\dots,N$. The scaling equation for the Hamiltonian Eq.~(\ref{cs0}) has the form \cite{hewson}
\begin{eqnarray}
\label{is}
\frac{dJ}{d\ln\Lambda}=-N J^2+N J^3.
\end{eqnarray}
For $N=2$ the model coincides with the spin-isotropic Kondo model. While Eq.~(\ref{is}) suggests the existence of a nontrivial fixed point $J=1$, we emphasize again that the equation becomes unreliable in the strong-coupling regime.

The anisotropic Kondo model represented in terms of Hubbard $X$-operators Eq.~(\ref{huhu}) has motivated one of us\cite{kogan2,kogan3} to consider the algebraic renormalizability of the anisotropic generalization of the CS model (or the ``$XYZ$ CS model'') for arbitrary $N$, Eq.~(\ref{cs01b}). This model proves to be algebraically renormalizable, and the scaling equations are\footnote{The second-order terms were obtained in the previous publication of one of the authors [Eq.~(46) of Ref.~\onlinecite{kogan2}], but due to an elementary algebra mistake, the analog of Eq.~(\ref{calingsc01c}) (and its special case for $N=3$ in Ref.~\onlinecite{kogan3}) contained an error.}
\begin{eqnarray}
\label{calingsc01c}
\frac{dJ_S}{d\ln\Lambda}&=&-(N-2)J_S^2-2J_SJ_z\nonumber\\
&&+J_S [(N-1)J_S^2+(N-3) J_A^2+J_z^2],\nonumber\\
%\label{scalingsc02b}
\frac{dJ_A}{d\ln\Lambda}&=&(N-2)J_A^2+2 J_AJ_z\\
&&+J_A [(N-3)J_S^2+(N-1) J_A^2+J_z^2],\nonumber\\
%\label{scalingsc03b}
\frac{dJ_z}{d\ln\Lambda}&=&-N (J_S^2-J_A^2)+N J_z(J_S^2+J_A^2). \nonumber
\end{eqnarray}
The second-order terms are derived explicitly in Appendix~\ref{sec:renorm}. Notice the symmetry of the Eq.~(\ref{calingsc01c}) with respect to simultaneous transformations
$J_S\leftrightarrow J_A$ and $\ln\Lambda \rightarrow -\ln\Lambda$. The $XXZ$ CS model introduced in Ref.~\onlinecite{kogan2} is obtained by setting $J_A=0$ (i.e. $J_x=J_y$) in Eqs.~(\ref{cs01b}) and (\ref{calingsc01c}). In the fully isotropic case $J_x=J_y=J_z$ from Eq.~(\ref{calingsc01c}) we recover the scaling equation for the original CS model Eq.~(\ref{is}).

Lacking analytical integration of Eq.~(\ref{calingsc01c}) for $N>2$ (even when the third-order terms are neglected), we concentrate on quantitative analysis of the flow diagram and of the phase diagram. There are three special particular solutions of the equation, corresponding to straight lines
\begin{eqnarray}
\label{s1}
&&J_A=0,\; J_S=J_z,\;\;\;\frac{dJ_z}{d\ln\Lambda}=-NJ_z^2+NJ_z^3; \\
\label{s2}
&&J_S=0,\;J_A=J_z,\;\;\;\frac{dJ_z}{d\ln\Lambda}=NJ_z^2+NJ_z^3; \\
\label{s5}
&&J_z=0,\;\;J_A=-J_S,\;\;\; \nonumber\\
&&\frac{dJ_S}{d\ln\Lambda} =-(N-2) J_S^2+2(N-2) J_S^3.
\end{eqnarray}
In addition, if we neglect the third-order terms, two more particular solutions emerge:
\begin{eqnarray}
\label{s3}
&&J_A=0,\;\; J_S=-\frac{2}{N}J_z,\;\;\;\frac{dJ_z}{d\ln\Lambda}=-\frac{4}{N}J_z^2;\\
\label{s4}
&&J_S=0,\;\;J_A=-\frac{2}{N}J_z,\;\;\;\frac{dJ_z}{d\ln\Lambda}=\frac{4}{N}J_z^2.
\end{eqnarray}
Consider, for example, Eq.~(\ref{s1}). The solution which starts at any $J_z>0$ blows up at a finite value of $\ln\Lambda$.
In addition, as shown in Appendix~\ref{sec:pfaff},  the ray $J_x=J_y=J_z>0$ is the attractor for the flow lines for arbitrary $N$.
We identify the appropriate strong-coupling phase with the attraction region  of the solution Eq.~(\ref{s1}).
The same is relevant for other three rays defined by Eqs.~(\ref{s2}), (\ref{s3}) and (\ref{s4}).

Additional support for this assumption comes from the analysis of the separatrices. From Eq.~(\ref{calingsc01c}) it follows that, similar to the case of Kondo model, the flow diagram has three invariant planes:
\begin{equation}
\label{pl6_1}
1)\;J_S=0,\;\;\;2)\;J_A=0,\;\;\;3)\;J_S+J_A=J_z.
\end{equation}
In addition, when we neglect the third-order terms, there is another set of three invariant planes,
\begin{eqnarray}
\label{pl6_2}
&&4)\;\frac{N}{2}(J_S+J_A)=-J_z\nonumber\\
&&5)\;J_S-\frac{N}{2} J_A=J_z,\\
&&6)\;\frac{N}{2}J_S-J_A=-J_z.\nonumber
\end{eqnarray}
One can easily see that for $N=2$  these are the invariant planes given by Eq.~(\ref{sp}).

Like it was in the case of Kondo model, parts of these invariant planes are separatrices.
Thus, for example Eq.~(\ref{calingsc01c}) leads to
\begin{eqnarray}
&&\frac{d}{d\ln\Lambda}\left(\frac{N}{2} J_S-J_A+J_z\right) \nonumber\\
=&&-\left(N J_S+2J_A\right)\left(\frac{N}{2} J_S-J_A+J_z\right) +O(J^3).
\end{eqnarray}
Hence when $N J_S+2J_A>0$, the last plane from Eq.~(\ref{pl6_2}) is an approximate phase boundary since any infinitesimal deviation from it is a relevant perturbation.
A similar analysis can be performed for other invariant planes.

Let us analyze the flow lines on the invariant planes.
Though taking $J_A=0$ or $J_S=0$ no longer brings us back to the familiar equations for the anisotropic Kondo model, the KT physics remains in action; it is only the separatrix lines that are different from before \cite{kogan2}. Notice that half of each line given by Eqs.~(\ref{s1}) - (\ref{s4}) serves as an attractor, and the other half as a separatrix line. For instance, provided $J_A=0$, systems whose initial parameters are located between the two separatrix lines [i.e. $J_z<0$, $J_z<J_S<-(2/N)J_z$] will flow to one of the $J_S=0$, $J_z<0$ fixed points, and other systems flow to strong coupling $J_S\to \pm \infty$, $J_z\to \infty$ depending on the initial sign of $J_S$.

Consider now the fixed points of the scaling equation. Similar to the spin-anisotropic Kondo model for $N=2$,  Eq.~(\ref{calingsc01c}) indicates that the $XYZ$ CS model has a line of fixed points $J_S=J_A=0$ for any $N$. Since all $N$-dependent terms are quadratic in $J_S$ or $J_A$, linearization again tells us that these fixed points are semistable, with $|J_S|$ relevant and $J_A$ irrelevant for $J_z>0$, and $|J_A|$ relevant and $J_S$ irrelevant for $J_z<0$.
However, the Kondo model  line of fixed points $J_S=J_A$, $J_z=0$ for $N=2$ is replaced by the line $J_z=-(N/2-1)J_S=-(N/2-1)J_A$, which does not lie on the $J_S$-$J_A$ plane.
On the other hand,  the line (\ref{s5}) for $N>2$ is not a fixed line like it was for Kondo model, but describes two flow trajectories instead: $J_S<0$ flows to  the origin, and  $J_S>0$ flows to strong coupling.

\begin{figure}
\begin{center}
\includegraphics[width=0.8\columnwidth]{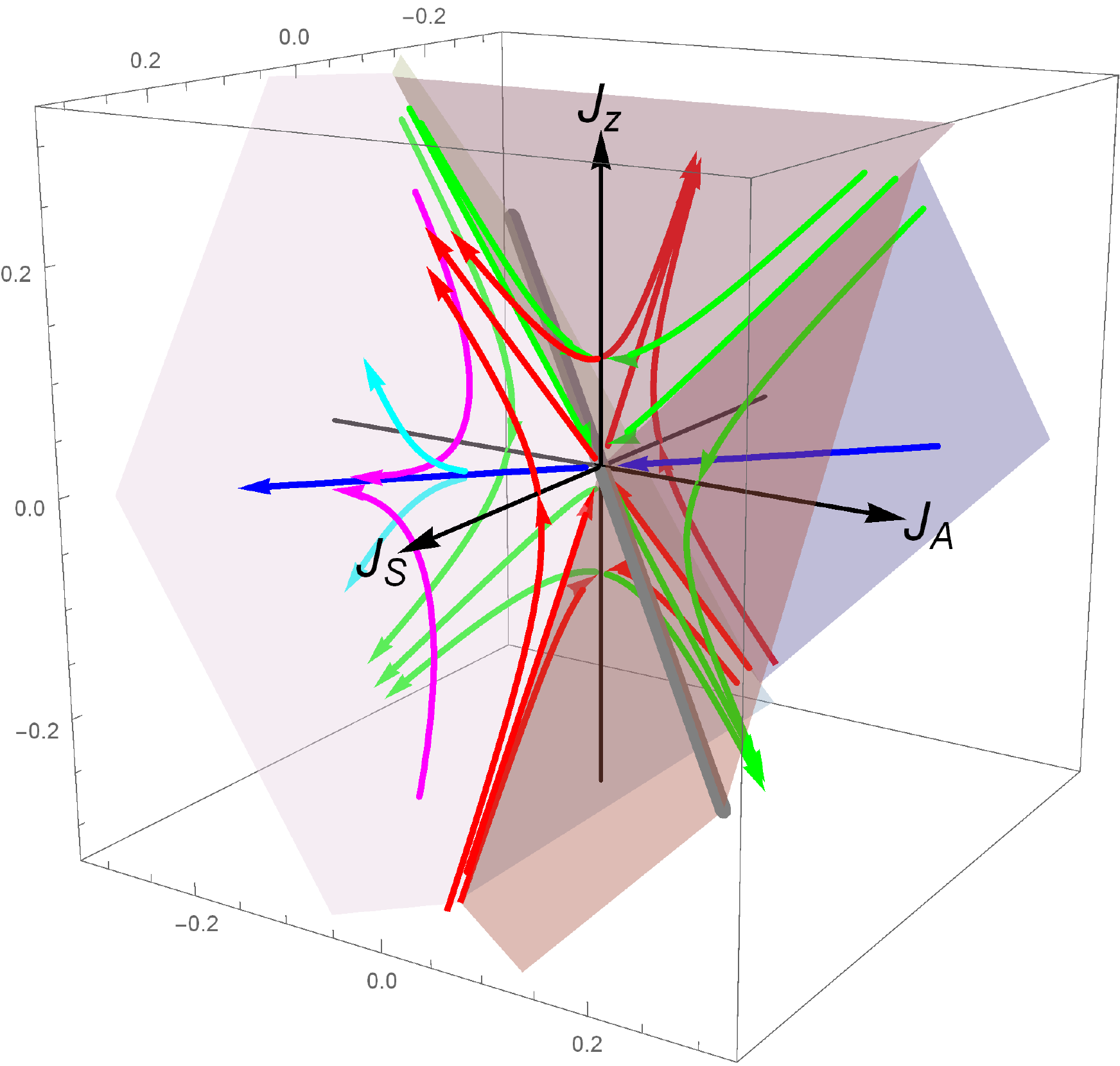}
\end{center}
\caption{Numerical three-dimensional flow diagram of the $XYZ$ CS model with $N=3$, calculated from the third-order weak-coupling scaling equation Eq.~(\ref{calingsc01c}). The symmetries $J_S\to -J_S$ and $J_A\to-J_A$ are now lost in contrast to the $N=2$ Kondo case. The red trajectories lie on the $J_A=0$ plane and the green ones lie on the $J_S=0$ plane; the KT separatrix lines are no longer reflection-symmetric with respect to the $J_S=0$ or $J_A=0$ planes. The gray line is the approximate fixed line $J_z=-(N/2-1)J_S=-(N/2-1)J_A$ near which the flow is third-order, and the blue trajectories satisfy $J_S=-J_A$, $J_z=0$. The remaining colored curves are typical trajectories which reside on one of the phase boundaries given in the text (magenta) or flow away from that phase boundary (cyan). All four strong-coupling phases are shown: $J_S\to \pm \infty$ and $J_z\to\infty$; $J_A\to \pm \infty$ and $J_z\to-\infty$.}\label{fig:flowCS}
\end{figure}

%{\color{red}Due to the fact that Eq.~(\ref{calingsc01c}) is scale invariant, all the geometric notions connected with its solutions (invariant planes, separatrices, attractors, separatrix lines) should also be scale invariant. That is they should be either rays (straight lines) or planes. Hence the geometric analysis in this Section is complete. In particular,}
We illustrate our findings of this Section in the flow diagram Fig.~\ref{fig:flowCS} for the $N=3$ case. As shown in the figure, the four strong-coupling phases mentioned above exhaust the phase diagram (apart from the critical points) in the weak-coupling regime, where our perturbative scaling method is applicable.

\subsection{Pseudogap density of states}

\begin{figure}
\begin{center}
\includegraphics[width=0.9\columnwidth]{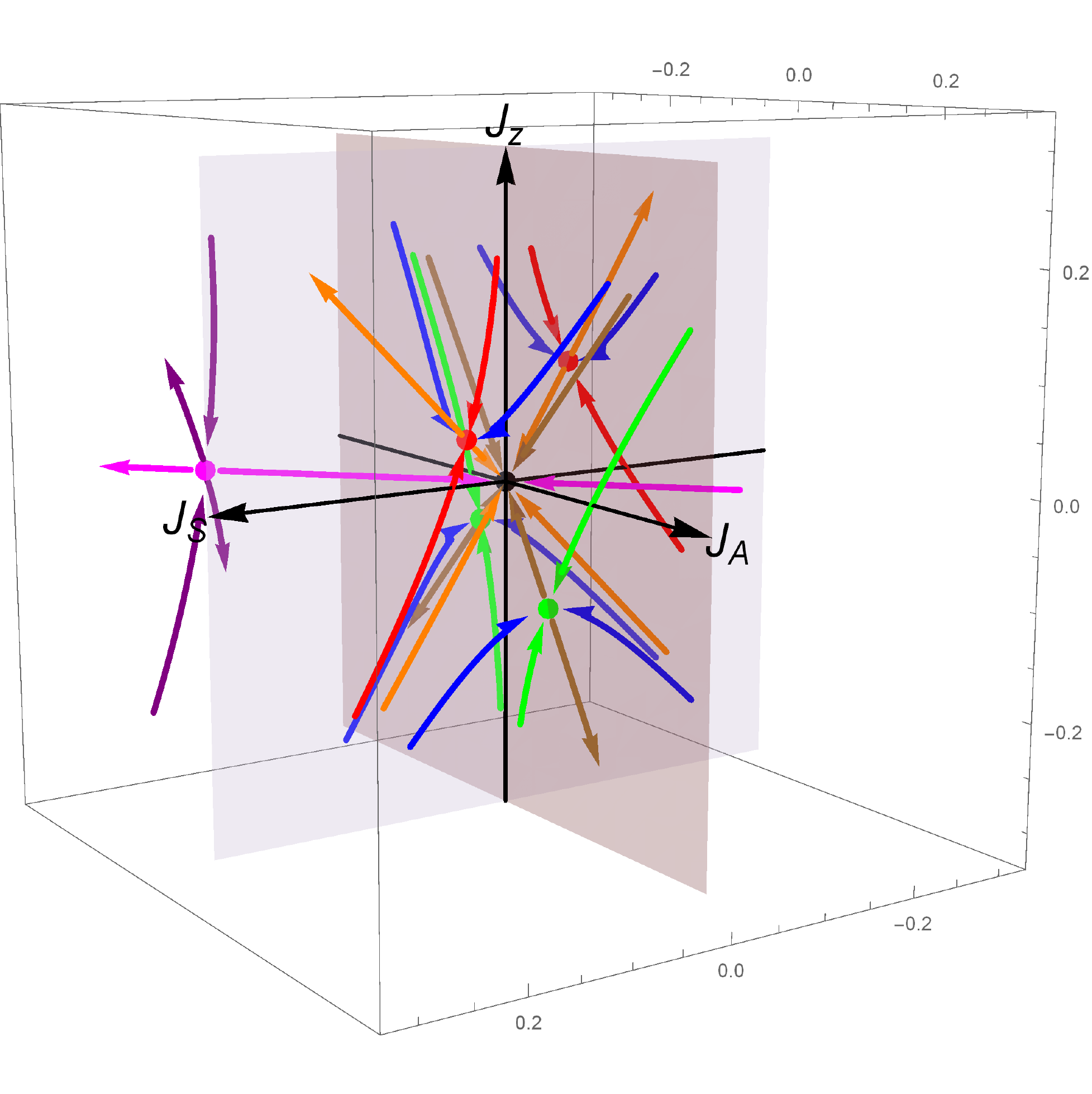}
\end{center}
\caption{Numerical three-dimensional flow diagram of the pseudogap $XYZ$ CS model with $N=3$ and $r=0.12$, calculated from the third-order weak-coupling scaling equation Eq.~(\ref{calingsc01p}). There are no longer any fixed lines in contrast to the constant-density-of-states model, but all trajectories connecting the trivial fixed point (black) with the nontrivial fixed points (red, green and magenta) are straight lines when $r$ is small. The red and orange trajectories lie on the $J_A=0$ plane, the green and brown ones lie on the $J_S=0$ plane, and the blue trajectories represent perturbations away from these planes. The purple trajectories represent perturbations away from the magenta fixed point with $J_z=0$. The strong-coupling phases are identical to those in the case of a constant density of states.}\label{fig:flowCSp}
\end{figure}

When the density of states takes a power-law form Eq.~(\ref{e}), the scaling equation for the $XYZ$ CS model becomes
\begin{eqnarray}
\label{calingsc01p}
\frac{dJ_S}{d\ln\Lambda}&=&r J_S-(N-2)J_S^2-2J_SJ_z\nonumber\\
&&+J_S [(N-1)J_S^2+(N-3) J_A^2+J_z^2],\nonumber\\
%\label{scalingsc02b}
\frac{dJ_A}{d\ln\Lambda}&=&r J_A+(N-2)J_A^2+2 J_AJ_z\\
&&+J_A [(N-3)J_S^2+(N-1) J_A^2+J_z^2],\nonumber\\
%\label{scalingsc03b}
\frac{dJ_z}{d\ln\Lambda}&=&r J_z-N (J_S^2-J_A^2)+N J_z(J_S^2+J_A^2).\nonumber
\end{eqnarray}
We plot the corresponding flow diagram in Fig.~\ref{fig:flowCSp}.

There is first and foremost a trivial decoupled fixed point $J_S=J_A=J_z=0$ (black). In addition, several nontrivial fixed points $(J_S,J_A,J_z)$ exist where one of the coupling constants vanishes:
\begin{eqnarray}
\frac{1}{2}(1-\sqrt{1-\frac{4r}{N}})(1,0,1)\;\;\text{(red)},\nonumber\\
\frac{1}{2}(1-\sqrt{1-\frac{4r}{N}})(0,-1,-1)\;\;\text{(green)},\nonumber\\
\frac{1}{2}(1-\sqrt{1-\frac{4r}{N-2}})(1,-1,0)\;\;\text{(magenta)},\\
(-r/2,0,Nr/4)+O(r^2)\;\;\text{(red)},\nonumber\\
(0,r/2,-Nr/4)+O(r^2)\;\;\text{(green)}.\nonumber
\end{eqnarray}
As before, in order for the perturbative treatment to be valid, these fixed points are only meaningful for small $r$. For the last two fixed points we have only kept the lowest-order terms in $r$; the full expressions are lengthy as they require solving quartic equations.

The four $J_A=0$ and $J_S=0$ fixed points are critical points with two stable directions, one in-plane and the other out-of-plane. Therefore, these four fixed points belong to the phase boundaries which separate the strong-coupling phases from the decoupled phase; these phase boundaries intersect the $J_A=0$ ($J_S=0$) plane at the red (green) trajectories, which are themselves separatrix lines on the coordinate planes. It is plausible that the $J_z=0$ nontrivial fixed point, which has only one stable direction, sits exactly at the intersection of two such phase boundaries.

As shown in Fig.~\ref{fig:flowCSp}, all of the following straight lines between the trivial fixed point and the nontrivial ones are valid scaling trajectories: $J_S=J_z, J_A=0$ and $J_z=-NJ_S/2, J_A=0$ (orange), $J_A=J_z, J_S=0$ and $J_z=-NJ_A/2, J_S=0$ (brown) and $J_S=-J_A, J_z=0$ (magenta). Each of these lines is divided into three segments by the trivial, stable fixed point and the nontrivial, unstable fixed point.

Based on our observations, as with the $N=2$ case, we expect the phase space for the $N>2$ pseudogap $XYZ$ CS model to be divided into five phases, namely a weak-coupling phase controlled by the trivial fixed point $J_S=J_A=J_z=0$ and four strong-coupling phases $J_S\to \pm \infty, J_z\to \infty$ and $J_A\to \pm \infty, J_z\to -\infty$. Unfortunately, the important linear terms in Eq.~(\ref{calingsc01p}) make it a difficult task in general to determine the phase boundaries or scaling invariants analytically.

\section{Conclusions\label{sec:conclusion}}

In this work, we have derived the third-order poor man's scaling equation for a quantum impurity model in an itinerant electron gas in the weak-coupling regime. Our theory is applied to the $XYZ$ Coqblin--Schrieffer model which was introduced by one of us earlier and is shown to be algebraically renormalizable. We write down the poor man's scaling equations under constant and pseudogap densities of states, and discuss their solutions for both the $N=2$ case (the anisotropic $XYZ$ Kondo model) and the $N>2$ case in detail. The corresponding three-dimensional weak-coupling flow diagrams are presented.

\begin{acknowledgments}

The results presented in this paper were obtained during E.K.'s visit to Max-Planck-Institut f\"{u}r Physik komplexer Systeme in December of 2019 and January of 2020. E.K. cordially thanks the Institute for the hospitality extended to him during that and all his previous visits. The authors are grateful to  V. Yu. Irkhin and P. Zalom  for valuable discussions.

\end{acknowledgments}

\begin{appendix}

\section{Third-order scaling equation Eq.~\ref{v3sceq}\label{sec:appv3}}

To obtain the third-order terms in the scaling equation Eq.~\ref{v3sceq}, following Ref.~\onlinecite{hewson}, Appendix~D, we should take into account the energy dependence of the effective Hamiltonian which is neglected at the second order. For the generic impurity Hamiltonian Eq.~\ref{hamilto}, when reducing the semi-bandwidth from $D$ to $D-\left\vert \delta D\right\vert $, the $O\left( V^{2}\right) $ correction to the low-energy Hamiltonian as represented by Fig.~\ref{fig:feynman} (a) is written as

\begin{align}
& \rho \left\vert \delta D\right\vert \sum_{k^{\prime }q}\sum_{\alpha \beta
ab}\sum_{\alpha ^{\prime }a^{\prime }}\frac{V_{\beta \alpha ,ba}V_{\alpha \alpha
^{\prime },aa^{\prime }}X_{ba^{\prime }}c_{k^{\prime }\beta }^{\dag
}c_{q\alpha ^{\prime }}}{E-D+\epsilon _{q}}  \notag \\
& +\rho \left\vert \delta D\right\vert \sum_{kq^{\prime }}\sum_{\alpha \beta
ab}\sum_{\beta ^{\prime }a^{\prime }}\frac{V_{\beta \alpha ,ba}V_{\beta ^{\prime
}\beta ,aa^{\prime }}X_{ba^{\prime }}c_{k\alpha }c_{q^{\prime }\beta
^{\prime }}^{\dag }}{E-D-\epsilon _{q^{\prime }}}\text{.}
\end{align}%
Tracing out the conduction electrons, we find the $O\left( V^{2}\right) $ correction to the impurity part of the effective Hamiltonian:

\begin{align}
& \rho ^{2}\left\vert \delta D\right\vert \sum_{\alpha \beta
ab}\sum_{a^{\prime }}V_{\beta \alpha ,ba}V_{\alpha \beta ,aa^{\prime
}}X_{ba^{\prime }} \notag \\
& \times\left[ \int_{-D+\left\vert \delta D\right\vert
}^{0}\frac{d\epsilon _{q}}{E-D+\epsilon _{q}}+\int_{0}^{D-\left\vert \delta D\right\vert }\frac{d\epsilon _{k}}{
E-D-\epsilon _{k}}\right]   \notag \\
& =\rho ^{2}\left\vert \delta D\right\vert \sum_{ab}\sum_{\alpha \beta
c}V_{\beta \alpha ,bc}V_{\alpha \beta ,ca}X_{ba}\left( -2\ln 2-\frac{E}{D}%
\right) \text{.}
\end{align}%
The term linear in energy $E$ will play an especially important role in the following:

\begin{equation}
-E\frac{\left\vert \delta D\right\vert }{D}\rho ^{2}\sum_{ab}\sum_{\alpha
\beta c}V_{\beta \alpha ,bc}V_{\alpha \beta ,ca}X_{ba}\text{.}
\end{equation}%
We emphasize that this is an operator in the impurity Hilbert space.

We now turn to the third-order diagram in Fig.~\ref{fig:feynman} (b). Contracting the fermion lines, this $O\left( V^{3}\right) $ diagram reads

\begin{eqnarray}
&&\sum_{kq}\sum_{\alpha \beta ab}\sum_{k_{1}k_{1}^{\prime }}\sum_{\alpha
_{1}\beta _{1}}\sum_{a^{\prime }b^{\prime }}c_{q\beta }^{\dag }c_{k\alpha }c_{k_{1}^{\prime }\beta _{1}}^{\dag
}c_{k_{1}\alpha _{1}}c_{k\alpha }^{\dag }c_{q\beta }\notag \\
&&\times \frac{V_{\beta \alpha ,ba}V_{\beta
_{1}\alpha _{1},ab^{\prime }}V_{\alpha \beta ,b^{\prime }a^{\prime
}}X_{ba^{\prime }}}{\left(E-\epsilon _{k}+\epsilon _{q}-\epsilon _{k_{1}^{\prime
}}+\epsilon _{k_{1}}\right) \left( E-\epsilon _{k}+\epsilon _{q}\right)}  \text{;}
\end{eqnarray}%
here we should contract $c_{k\alpha }$ with $c_{k\alpha }^{\dag }$ and $c_{q\beta }^{\dag }$ with $c_{q\beta }$. If the virtual particle $c_{k\alpha}^{\dag }$ resides in the energy range to be integrated out (i.e. $\epsilon _{k}\approx D$), this gives a contribution

\begin{align}
& \rho ^{2}\left\vert \delta D\right\vert \sum_{\alpha \beta
ab}\sum_{k_{1}k_{1}^{\prime }}\sum_{\alpha _{1}\beta _{1}}\sum_{a^{\prime
}b^{\prime }}V_{\beta \alpha ,ba}V_{\beta _{1}\alpha _{1},ab^{\prime
}}V_{\alpha \beta ,b^{\prime }a^{\prime }}  \notag \\
&\times \int_{-D+\left\vert \delta
D\right\vert }^{0}\frac{d\epsilon _{q} X_{ba^{\prime }} c_{k_{1}^{\prime
}\beta _{1}}^{\dag }c_{k_{1}\alpha _{1}}}{\left(E-D+\epsilon _{q}-\epsilon
_{k_{1}^{\prime }}+\epsilon _{k_{1}}\right) \left( E-D+\epsilon _{q}\right)}  \notag \\
& \approx\rho ^{2}\frac{\left\vert \delta D\right\vert }{2D}\sum_{\alpha \beta
ab}\sum_{k_{1}k_{1}^{\prime }}\sum_{\alpha _{1}\beta _{1}}\sum_{a^{\prime
}b^{\prime }}V_{\beta \alpha ,ba}V_{\beta _{1}\alpha _{1},ab^{\prime
}}V_{\alpha \beta ,b^{\prime }a^{\prime }}   \notag \\
& \times X_{ba^{\prime }}c_{k_{1}^{\prime
}\beta _{1}}^{\dag }c_{k_{1}\alpha _{1}}\text{;}
\end{align}%
on the other hand, if $\epsilon _{q}\approx -D$, we have the virtual hole contribution

\begin{align}
& \rho ^{2}\left\vert \delta D\right\vert \sum_{\alpha \beta
ab}\sum_{k_{1}k_{1}^{\prime }}\sum_{\alpha _{1}\beta _{1}}\sum_{a^{\prime
}b^{\prime }}V_{\beta \alpha ,ba}V_{\beta _{1}\alpha _{1},ab^{\prime
}}V_{\alpha \beta ,b^{\prime }a^{\prime }}  \notag \\
&\times \int_{0}^{D-\left\vert \delta
D\right\vert }\frac{d\epsilon _{k}X_{ba^{\prime }}c_{k_{1}^{\prime
}\beta _{1}}^{\dag }c_{k_{1}\alpha _{1}}}{\left(E-D-\epsilon _{k}-\epsilon
_{k_{1}^{\prime }}+\epsilon _{k_{1}}\right) \left( E-D-\epsilon _{k}\right) }  \notag \\
& \approx\rho ^{2}\frac{\left\vert \delta D\right\vert }{2D}\sum_{\alpha \beta
ab}\sum_{k_{1}k_{1}^{\prime }}\sum_{\alpha _{1}\beta _{1}}\sum_{a^{\prime
}b^{\prime }}V_{\beta \alpha ,ba}V_{\beta _{1}\alpha _{1},ab^{\prime
}}V_{\alpha \beta ,b^{\prime }a^{\prime }}\notag \\
& \times X_{ba^{\prime }}c_{k_{1}^{\prime
}\beta _{1}}^{\dag }c_{k_{1}\alpha _{1}}\text{.}
\end{align}%
The virtual particle and hole contributions are thus identical.

To find the total third-order scaling contribution to the coupling constant, we write the effective Hamiltonian as

\begin{align}
H_{\text{eff}}\left( E\right) &=\left( 1+S\right) ^{-\frac{1}{2}}H_{\text{eff}%
}\left( 0\right) \left( 1+S\right) ^{-\frac{1}{2}}  \notag \\
&\approx \left( 1-\frac{1}{2%
}S\right) H_{\text{eff}}\left( 0\right) \left( 1-\frac{1}{2}S\right) \text{,}
\end{align}%
where the effective impurity Hamiltonian at $E=0$ is, to $O\left( V^{3}\right) $,

\begin{eqnarray}
&&H_{\text{eff}}\left( 0\right)  =\sum_{\mathbf{kk}^{\prime }}\sum_{\alpha
\beta ab}[ V_{\beta \alpha ,ba}-\rho \frac{\left\vert \delta
D\right\vert }{D} \sum_{\gamma c} \notag \\
&&\times \left( V_{\beta \gamma ,bc}V_{\gamma \alpha
,ca}-V_{\gamma \alpha ,bc}V_{\beta \gamma ,ca}\right) +\rho ^{2}\frac{\left\vert \delta D\right\vert }{D} \notag \\
&& \times \sum_{\delta
\gamma }\sum_{cd}V_{\delta \gamma ,bc}V_{\beta \alpha ,cd}V_{\gamma \delta
,da}] X_{ba}c_{\mathbf{k}^{\prime }\beta }^{\dag }c_{\mathbf{k}\alpha }%
\text{,}
\end{eqnarray}%
and the wave function renormalization $S\equiv -\partial H_{\text{eff}}\left(
E\right) /\partial E$ is to $O\left( V^{2}\right) $

\begin{equation}
S=\rho ^{2}\frac{\left\vert \delta D\right\vert }{D}\sum_{\alpha \beta
}\sum_{abc}V_{\beta \alpha ,bc}V_{\alpha \beta ,ca}X_{ba}\text{.}
\end{equation}%
Putting everything together, we obtain Eq.~\ref{v3sceq}.

\section{Algebraic renormalizability of the $XYZ$ CS model\label{sec:renorm}}

The  commutation relations necessary for writing down scaling equation for the interaction Eq.~(\ref{cs01b}) are
\begin{eqnarray}
%\label{cs01b}
[c_{m'}^{\dagger}c_{m},c_{m'''}^{\dagger}c_{m''}]=\delta_{mm'''}c_{m'}^{\dagger}c_{m''}-\delta_{m'm''}c_{m'''}^{\dagger}c_{m}.
\end{eqnarray}
(and the same for the Hubbard operators).
Substituting the terms in the R.H.S. of Eq.~(\ref{cs01b}) into the second-order term in Eq.~(\ref{v3sceq}) we get:

\noindent
for the $J_S^2$ terms
\begin{eqnarray}
\label{oducti}
&&\sum_{m\neq m'}\sum_{m''\neq m'''}[X_{mm'},X_{m''m'''}] [c_{m'}^{\dagger}c_{m},c_{m'''}^{\dagger}c_{m''}]\nonumber\\
&=&\sum_{m\neq m'}\sum_{m''\neq m'''}\left(X_{mm'''}\delta_{m'm''}-X_{m''m'}\delta_{mm'''}\right)\nonumber\\
&\cdot&\left(c_{m'}^{\dagger}c_{m''}\delta_{mm'''}-c_{m'''}^{\dagger}c_{m}\delta_{m'm''}\right)\nonumber\\
&=&2\left\{-(N-2)\sum_{m\neq m'''}X_{mm'''}c_{m'''}^{\dagger}c_{m}\right.\nonumber\\
&-&\left.(N-1)\sum_{m}X_{mm}c_{m}^{\dagger}c_{m}+\sum_{m\neq m'}X_{mm}c_{m'}^{\dagger}c_{m'}\right\}\nonumber\\
&=&2\left\{-(N-2)\sum_{m\neq m'''}X_{mm'''}c_{m'''}^{\dagger}c_{m}\right.\nonumber\\
&-&\left.N\sum_{m}X_{mm}c_{m}^{\dagger}c_{m}+\sum_{mm'}X_{mm}c_{m'}^{\dagger}c_{m'}\right\};
\end{eqnarray}

\noindent
for the $J_A^2$ terms
\begin{eqnarray}
\label{cti}
&&\sum_{m\neq m'}\sum_{m''\neq m'''}[X_{mm'},X_{m''m'''}] [c_{m}^{\dagger}c_{m'},c_{m''}^{\dagger}c_{m'''}]\nonumber\\
&=&\sum_{m\neq m'}\sum_{m''\neq m'''}\left(X_{mm'''}\delta_{m'm''}-X_{m''m'}\delta_{mm'''}\right)\nonumber\\
&\cdot&\left(c_{m}^{\dagger}c_{m'''}\delta_{m'm''}-c_{m''}^{\dagger}c_{m'}\delta_{mm'''}\right)\nonumber\\
&=&2\left\{(N-2)\sum_{m\neq m'''}X_{mm'''}c_{m}^{\dagger}c_{m'''}\right.\nonumber\\
&+&\left.(N-1)\sum_{m}X_{mm}c_{m}^{\dagger}c_{m}-\sum_{m\neq m'}X_{mm}c_{m'}^{\dagger}c_{m'}\right\}\nonumber\\
&=&2\left\{(N-2)\sum_{m\neq m'''}X_{mm'''}c_{m}^{\dagger}c_{m'''}\right.\nonumber\\
&+&\left.N\sum_{m}X_{mm}c_{m}^{\dagger}c_{m}-\sum_{mm'}X_{mm}c_{m'}^{\dagger}c_{m'}\right\};
\end{eqnarray}

\noindent
for the $J_SJ_z$ terms
\begin{eqnarray}
\label{ctib}
&&\sum_{m\neq m',m''}[X_{mm'},X_{m''m''}] [c_{m'}^{\dagger}c_{m},c_{m''}^{\dagger}c_{m''}]\nonumber\\
&=&\sum_{m\neq m',m''}\left(X_{mm''}\delta_{m'm''}-X_{m''m}\delta_{mm''}\right)\nonumber\\
&\cdot&\left(c_{m'}^{\dagger}c_{m''}\delta_{mm''}-c_{m''}^{\dagger}c_{m}\delta_{m'm''}\right)\nonumber\\
&=&-2\sum_{m\neq m'}X_{mm'}c_{m'}^{\dagger}c_{m};
\end{eqnarray}

\noindent
for the $J_AJ_z$ terms
\begin{eqnarray}
\label{tib}
&&\sum_{m\neq m',m''}[X_{mm'},X_{m''m''}] [c_{m}^{\dagger}c_{m'},c_{m''}^{\dagger}c_{m''}]\nonumber\\
&=&\sum_{m\neq m',m''}\left(X_{mm''}\delta_{m'm''}-X_{m''m}\delta_{mm''}\right)\nonumber\\
&\cdot&\left(c_{m}^{\dagger}c_{m''}\delta_{m'm''}-c_{m''}^{\dagger}c_{m'}\delta_{mm''}\right)\nonumber\\
&=&2\sum_{m\neq m'}X_{mm'}c_{m}^{\dagger}c_{m'}.
\end{eqnarray}
Equations (\ref{oducti}) - (\ref{tib}) combined together give us scaling equation (\ref{calingsc01c}).

\section{Attractors and separatrix lines\label{sec:pfaff}}

Scaling flows of Eq.~(\ref{calingsc01c}) in the vicinity of the ray described by Eq.~(\ref{s1}) can be presented as
\begin{eqnarray}
J_S&=&\left(1+c_1t^{\rho}\right)\frac{1}{Nt}\nonumber\\
J_A&=&c_2t^{\rho}\frac{1}{Nt}\nonumber\\
J_z&=&\left(1+c_3t^{\rho}\right)\frac{1}{Nt},
\end{eqnarray}
where $t=\ln\Lambda$; the solution blows up when $t\to +0$.
Linearizing with respect to small deviations  we obtain the matrix equation
\begin{eqnarray}
\label{rho}
KC=\rho C,
\end{eqnarray}
where $C = (c_1,c_2,c_3)^T$, and the Kovalevskaya matrix \cite{yoshida2,goriely} (KM) $K$ is
\begin{eqnarray}
\label{ii2}
K=\left(\begin{array}{ccc}\frac{2}{N}-1 & 0 &  -\frac{2}{N} \\
0 & \frac{2}{N}+1 & 0 \\ -2 & 0 & 1 \end{array}\right).
\end{eqnarray}
After elementary algebra we obtain the eigenvalues of the KM, which are called Kovalevskaya exponents (KEs): $\rho_1=-1$, and
doubly degenerate $\rho_2 = 2/N+1$. Thus we have two independent solutions to Eq.~(\ref{rho}).

The solution corresponding to $\rho_1$ is irrelevant; it has the form
$C=(1,0,1)^T$ and represents just the shift of the ray $J_S=J_z=1/Nt,J_A=0$ along itself.
The KE $\rho_2$ gives us two independent solutions, corresponding to
$C=(1,0,-1)^T$ and  $C=(0,1,0)^T$.
Thus we see that the ray described by Eq.~(\ref{s1}) is an attractor.
From the symmetry of Eq.~(\ref{calingsc01c}) it follows that the behavior of solutions in the vicinity of the ray Eq.~(\ref{s2}) is identical to that in the vicinity of the ray  Eq.~(\ref{s1}).

Calculation of the KEs can be formalized  \cite{yoshida2,goriely}. Consider a system of ordinary differential equations
\begin{eqnarray}
\label{pf1}
\frac{dx_i}{dt}=f_i(x_1,\dots,x_n).
\end{eqnarray}
Consider a solution $C = (c_1,\dots,c_n)^T\neq(0, \dots, 0)^T$ of algebraic
equations
\begin{eqnarray}
\label{pf2}
f_i(c_1,\dots,c_n)+c_i=0.
\end{eqnarray}
We define the KM
\begin{eqnarray}
\label{ii}
K_{ij}=\frac{\partial f_i}{\partial x_j}(C)+\delta_{ij}.
\end{eqnarray}
The eigenvalues $\rho_1,\dots,\rho_n$ of the KM are the KEs.

For Eq.~(\ref{calingsc01c})
\begin{eqnarray}
\label{ii2b}
&&K_{ij}=\\
&&\left(\begin{array}{ccc} 1-2(N-2)c_1 -2c_3 & 0 &  -2c_1 \\
0 & 1+2(N-2)c_2+2c_3 & 2c_2 \\ -2Nc_1 & 2Nc_2 & 1 \end{array}\right), \nonumber
\end{eqnarray}
where $c_1,c_2,c_3$ are the solutions of the equation
\begin{eqnarray}
\label{str}
-(N-2)c_1^2-2c_1c_3+c_1&=&0 \nonumber\\
(N-2)c_2^2+2c_2c_3+c_2&=&0 \\
-N(c_1^2-c_2^2)+c_3&=&0.\nonumber
\end{eqnarray}
The ray Eq.~(\ref{s1}) corresponds to the solution of Eq.~(\ref{str}): $C=(1/N,0,1/N)$.
Substituting the solution into Eq.~(\ref{ii2b}) we recover Eq.~(\ref{ii2}).

The ray Eq.~(\ref{s3}) corresponds to the  solution of Eq.~(\ref{str}): $C=(-1/2,0,N/4)$.
Substituting the solution into Eq.~(\ref{ii2}) we obtain
\begin{eqnarray}
\label{ii2c}
K_{ij}=\left(\begin{array}{ccc} \frac{N}{2}-1 & 0 &  1 \\
0 & \frac{N}{2}+1 & 0 \\ N & 0 & 1 \end{array}\right).
\end{eqnarray}
After elementary algebra we obtain the KEs: $\rho_1=-1$, and doubly degenerate $\rho_2=N/2+1$.
As is always the case, the solution corresponding to $\rho_1$ is irrelevant; it has the form
$C=(1,0,N/2)^T$ and represents just the shift of the ray $J_S=-(2/N)J_z,J_A=0$ along itself.
The positive value of $\rho_2$ means that the ray in question is an attractor.
From the symmetry of Eq.~(\ref{calingsc01c}) it follows that the behavior of solutions in the vicinity of the ray Eq.~(\ref{s4}) is identical to that in the vicinity of the ray Eq.~(\ref{s3}).

The ray Eq.~(\ref{s5}) corresponds to the solution of Eq.~(\ref{str}): $C=(1/(N-2),-1/(N-2),0)$.
Substituting the solution into Eq.~(\ref{ii2b})
we obtain
\begin{eqnarray}
\label{ii4}
K=\left(\begin{array}{ccc} -1 & 0 &  -\frac{2}{N-2} \\
0 & -1 & -\frac{2}{N-2} \\ -\frac{2N}{N-2} & -\frac{2N}{N-2} & 1 \end{array}\right).
\end{eqnarray}
After elementary algebra we obtain the KEs: $\rho_1=-1$, $\rho_2=-(N+2)/(N-2)$ and $\rho_3=(N+2)/(N-2)$.
The solution corresponding to $\rho_1$  represents just the shift of the ray $J_S=-J_A,J_z=0$ along itself.
The second negative KE makes behavior of the flow line in the vicinity of the ray Eq.~(\ref{s5}) qualitatively different from that in the vicinity of the rays Eqs.~(\ref{s1}) - (\ref{s4}). It means that in general the flow lines diverge  from the
ray. However, the positive value of $\rho_3$ is the manifestation of the fact that in the plane $(N/2)(J_S+J_A)=-J_z$ the ray is an attractor.

Calculated KEs allow us to address the question of the existence of the first integrals for Eq.~(\ref{calingsc01c}).
Notice that the existence of two such independent integrals makes the analytical integration of Eq.~(\ref{scalingsc02}) possible. All
the KEs turn out to be rational, a necessary condition for the existence of polynomial first integrals of Eq.~(\ref{calingsc01c}) \cite{yoshida2,goriely}. Though the condition is not sufficient, one may hope that such integrals can be found.

\end{appendix}


\begin{thebibliography}{99}

\bibitem{kondo} J. Kondo,  Prog. Theor. Phys. {\bf 32}, 37 (1964).

\bibitem{hewson} A. C. Hewson, {\it The Kondo Problem to Heavy Fermions}, (Cambridge University Press, Cambridge, 1993).

\bibitem{anderson} P. W. Anderson, J. Phys. C {\bf 3}, 2436 (1970).

\bibitem{cox} D.L. Cox and A. Zawadowski, Adv. Phys. {\bf 47},  599 (1998).

\bibitem{coqblin} B. Coqblin, J.R. Schrieffer, Phys. Rev. {\bf 185}, 847 (1969).

\bibitem{kikoin} A. K. Kikoin, M. Kiselev, Y. Avishai, {\it Dynamical Symmetries
for Nanostructures} (Springer-Verlag/Wien, 2012).

\bibitem{desgranges} H.-U. Desgranges, Physica B: Condensed Matter {\bf 454}, 135 (2014); {\bf 473}, 93 (2015).

\bibitem{figueira} M.S.Figueira, A. Saguia, M.E.Foglio, J. Silva-Valencia, and R.Franco, Physica B: Condensed Matter
{\bf 455}, 92 (2014)

\bibitem{avishai}  I. Kuzmenko, T. Kuzmenko, Y. Avishai, and Gyu-Boong Jo, Phys. Rev. B  {\bf 93}, 115143 (2016);
ibid {\bf 97}, 075124, (2018).

\bibitem{costi} G. Zarand, T. Costi, A. Jerez, and N. Andrei, Phys. Rev. B {\bf 65}, 134416 (2002).

\bibitem{irkhin2} V.Yu. Irkhin, M.I. Katsnelson,
 Phys. Rev. B {\bf 59}, 9348 (1999).

\bibitem{thomas} C. Thomas,  A. S. da Rosa Simoes, C.  Lacroix, J.  R. Iglesias,  B. Coqblin,
Physica B, {\bf 404}, 3008 (2009).

\bibitem{shiba} H. Shiba, Prog.  Theor. Phys. {\bf 43}, 601 (1970).

\bibitem{kogan} E. Kogan, K. Noda, and S. Yunoki, Phys. Rev. B {\bf 95}, 165412 (2017).

\bibitem{yosida} K. Yosida, {\it Theory of Magnetism} (Springer, Berlin Heidelberg New York, 1996).

 \bibitem{kogan2} E. Kogan,  J. Phys. Commun. {\bf 2}, 085001 (2018).

\bibitem{kogan3} E. Kogan,  J. Phys. Commun. {\bf 3}, 125001 (2019).

\bibitem{fradkin} D. Withoff and E. Fradkin, \prl {\bf 64}, 1835 (1990).

\bibitem{fritzvojta} L. Fritz and M. Vojta, Phys. Rev. B {\bf 70}, 214427 (2004).

\bibitem{gbi} C. Gonzalez-Buxton and K. Ingersent, Phys. Rev. B {\bf 57}, 14254 (1998).

\bibitem{cassanello} C. R. Cassanello and E. Fradkin, Phys. Rev. B {\bf 53}, 15079 (1996).

\bibitem{vojtabulla} M. Vojta and R. Bulla, Eur. Phys. J. B {\bf 28}, 283 (2002).

\bibitem{mitchellfritz} A. K. Mitchell and L. Fritz, Phys. Rev. B {\bf 88}, 075104 (2013).

\bibitem{shinicaaffleck} Z. Shi. E. M. Nica and I. Affleck, Phys. Rev. B {\bf 100}, 125158 (2019).

%\bibitem{georgi} H. Georgi, {\it Lie Algebras in Particle Physics (2nd ed.)}, (Taylor \& Frencis, 2018).

%\bibitem{yoshida} H. Yoshida, Celestial Mech. {\bf 31}, 363 (1983).

\bibitem{yoshida2} H. Yoshida, B. Grammaticos, and A. Ramani, Acta Applicandae Mathematicae {\bf 8}, 75 (1987)

%\bibitem{tsy}  A. Tsygvintsev, J. Phys. A: Math. Gen. {\bf 34}, 2185 (2001).

\bibitem{goriely} A. Goriely, {\it Integrability and Nonintegrability of Dynamical Systems} (World Scientific Publishing Co. Pte. Ltd., Singapore, 2001).

\end{thebibliography}
\end{document}